\documentclass[11pt,twoside]{article}
\usepackage{asp2010}

\resetcounters

\begin{document}

\title{New Observational Evidence of Flash Mixing
on the White Dwarf Cooling Curve}
\author{T.M. Brown,$^1$ T. Lanz,$^2$ A.V. Sweigart,$^3$ Misty Cracraft,$^1$
Ivan Hubeny,$^4$ and W.B. Landsman$^5$
\affil{$^1$Space Telescope Science Institute, 3700 San Martin Drive,
Baltimore, MD 21218}
\affil{$^2$Department of Astronomy, University of Maryland, College
Park, MD 20742}
\affil{$^3$Code 667, NASA Goddard Space Flight Center, Greenbelt, MD
20771}
\affil{$^4$Steward Observatory, University of Arizona, Tucson,
  AZ 85712}
\affil{$^5$Adnet Systems, NASA Goddard Space
  Flight Center, Greenbelt, MD 20771
}
}

\begin{abstract}

Blue hook stars are a class of subluminous extreme horizontal branch
stars that were discovered in UV images of the massive globular
clusters $\omega$ Cen and NGC~2808.  These stars occupy a region of
the HR diagram that is unexplained by canonical stellar evolution
theory.  Using new theoretical evolutionary and atmospheric models, we
have shown that the blue hook stars are very likely the progeny of
stars that undergo extensive internal mixing during a late helium-core
flash on the white dwarf cooling curve.  This ``flash mixing''
produces hotter-than-normal EHB stars with atmospheres significantly
enhanced in helium and carbon.  The larger bolometric correction,
combined with the decrease in hydrogen opacity, makes these stars appear
subluminous in the optical and UV.  Flash mixing is more likely to
occur in stars born with a high helium abundance, due to their lower
mass at the main sequence turnoff.  For this reason, the phenomenon
is more common in those massive globular clusters that show evidence
for secondary populations enhanced in helium.  However, a high helium
abundance does not, by itself, explain the presence of blue hook stars
in massive globular clusters.  Here, we present new observational
evidence for flash mixing, using recent {\it HST} observations.  These
include UV color-magnitude diagrams of six massive globular clusters
and far-UV spectroscopy of hot subdwarfs in one of these clusters
(NGC~2808).

\end{abstract} 

\section{Introduction}

In an old stellar population of a given age and chemical composition,
the horizontal branch (HB) is the locus of stars undergoing stable
helium burning in the core, where the extent in effective temperature arises
from a range of envelope mass.  In globular clusters hosting extreme HB (EHB)
stars at temperatures $T_{\rm eff} \gtrsim 16,000$~K, 
the hot end of the HB appears nearly vertical in optical color-magnitude
diagrams (CMDs), due to a color degeneracy with temperature as the 
stellar luminosity shifts from optical to ultraviolet wavelengths.
However, UV photometry of $\omega$ Cen demonstrated that it hosts a
population of subluminous EHB stars that form a ``blue hook'' (BH) at
the hot end of the canonical EHB \citep{dcruz_1996,dcruz_2000}.
\citet{dcruz_1996} proposed that such stars may have undergone a
delayed helium-core flash after peeling away from the red giant branch (RGB) 
due to high mass loss.

\begin{figure}[t!]
\plotone{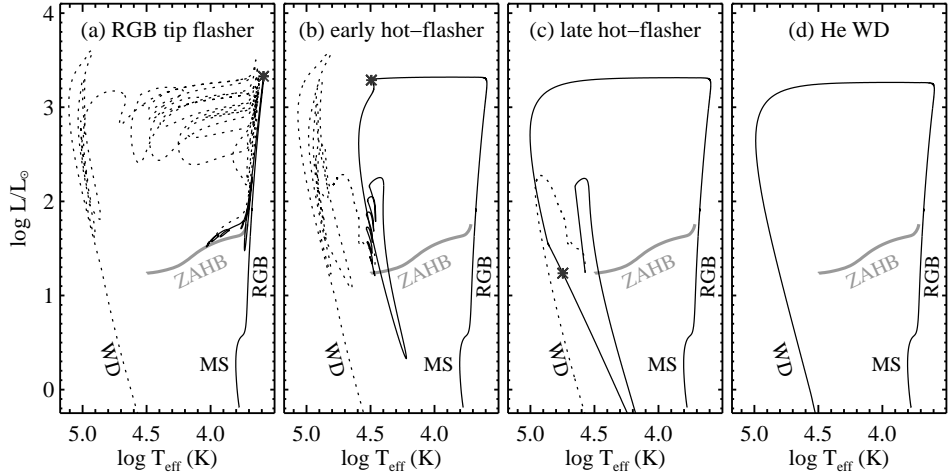}
\caption{ Various evolutionary paths for producing an HB star 
  \citep[from][]{brown_2010}. The zero-age HB (ZAHB) phase is highlighted in
  grey, while the pre-ZAHB evolution (solid curves) and post-ZAHB
  evolution (dashed curves) are in black. The peak of the helium-core
  flash is marked by an asterisk. The four panels show the evolution
  for progressively larger amounts of mass loss on the RGB. The
  evolution in the first two panels produces canonical HB stars in
  which the H-rich surface composition does not change during the
  helium-core flash. In the third panel, the helium-core flash occurs
  on the WD cooling curve, producing a flash-mixed star having a
  surface composition highly enriched in helium and carbon and a
  temperature significantly hotter than the canonical HB.  In the
  fourth panel, the helium flash never occurs and the star dies as a
  helium WD.}
\end{figure}

We now know that stars can follow a variety of evolutionary paths to
the HB, depending on when the helium-core flash occurs (see Figure 1).
For normal rates of mass loss, the helium-core flash will occur at the
RGB tip (RGB tip flasher, Figure 1a).  However, for sufficiently high
rates of mass loss, the helium-core flash can occur either during the
crossing of the HR diagram (early hot flasher, Figure 1b) or on the
white dwarf (WD) cooling curve (late hot flasher, Figure 1c), as first
shown by \citet{cast_1993}.  \citet{brown_2001} demonstrated that a
delayed helium-core flash on the WD cooling curve would result in a
stellar atmosphere extremely enhanced in helium ($\sim$96\% by mass),
with significant enhancements of carbon and/or nitrogen ($\sim$1 to
4\% by mass), due to the mixing of the stellar envelope into the hot
helium-burning core.  This ``flash mixing'' decreases the opacity
below the Lyman limit, thereby lowering the flux at longer
wavelengths, and increases the effective temperature in the EHB stars,
thereby increasing the bolometric correction.  \citet{brown_2001}
found that both of these effects together might explain the BH feature
in the UV CMD of NGC~2808.  Evidence for BH
populations in other clusters soon followed, including NGC~6715
\citep{rosenberg_2004}, NGC~6388 \citep{busso_2007}, NGC~2419
\citep{ripepi_2007}, NGC~6273 \citep[noted by][first
  appearance]{rosenberg_2004,piotto_1999}, and NGC~6441
\citep{busso_2007,dieball_1999}.

Here we present observational evidence for the flash-mixing mechanism
in those EHB stars that are subluminous with respect to the
expectations from canonical stellar evolution theory.  This evidence
includes new observations from the {\it Hubble Space Telescope (HST)}.
We will examine far-UV and near-UV photometry of massive globular
clusters spanning a wide range of metallicity, and far-UV spectroscopy
of both normal and subluminous EHB stars in one of these clusters
(NGC~2808).

\begin{figure}[t!]
\plotone{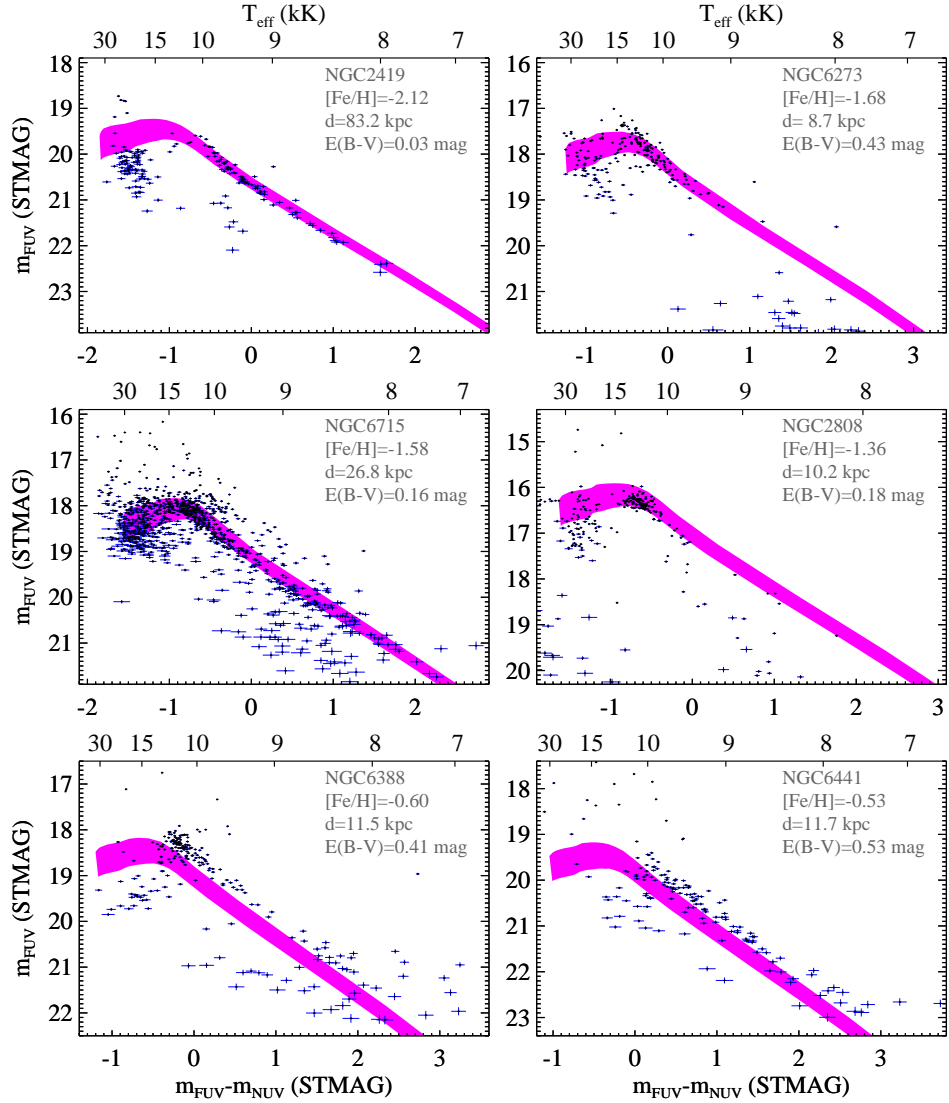}
\caption{
Ultraviolet CMDs for six massive globular clusters spanning a wide
range in [Fe/H] (points), along with the HB locus expected
from canonical stellar evolution theory at standard
helium abundance ($Y=0.23$; violet shaded area).  The assumed
cluster parameters are indicated (grey labels).
An approximate conversion (assuming normal cluster abundances) 
between observed color and effective temperature
is shown on the upper abscissa in each panel.  Photometric errors
(blue bars) are only significant for stars much redder and fainter
than the EHB stars that are the focus of this paper.
The BH population appears as a downward hook at the hot
end of the observed HB, where it deviates below the canonical EHB. 
For the two metal-rich clusters, that deviation occurs far to the red
of the hot end of the canonical HB.}
\end{figure}

\section{Imaging}

We have obtained far-UV and near-UV photometry of six massive globular
clusters \citep{brown_2001,brown_2010}.  The far-UV images were
obtained using the Space Telescope Imaging Spectrograph (STIS) and the
Solar Blind Channel on the Advanced Camera for Surveys (ACS).
The near-UV images were obtained using STIS, the Wide Field and
Planetary Camera 2, and the High Resolution Camera on
ACS.  Although the bandpasses employed vary from cluster to cluster,
the distinctions are not large enough to hamper comparisons
of the resulting CMDs, and are in any case included in our
transformations of models to the observational plane \citep[for details, see]
[]{brown_2010}.

\begin{figure}[t!]
\plotone{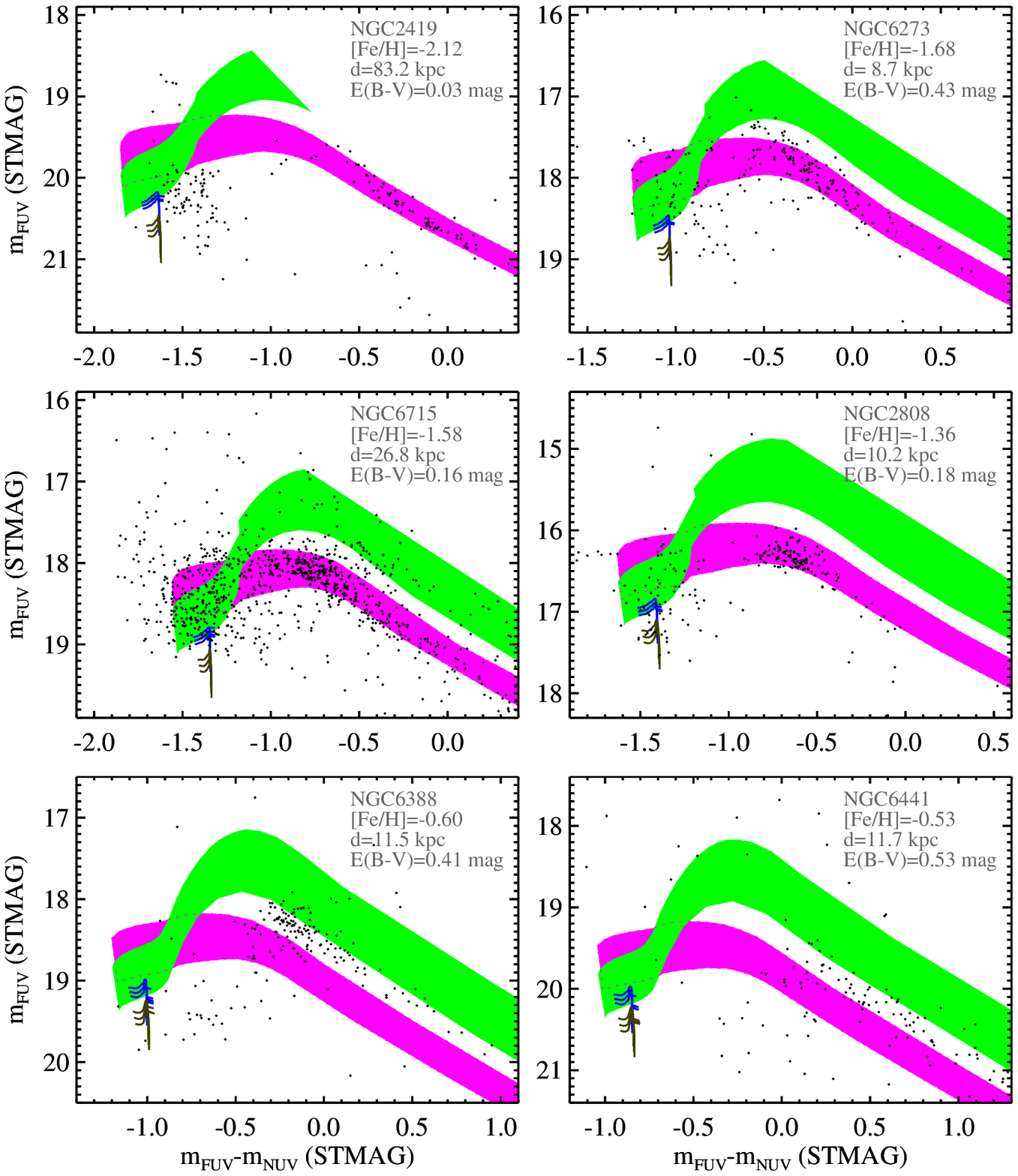}
\caption{Our UV CMDs compared to canonical models at standard helium
abundance ($Y=0.23$; violet), canonical models at enhanced helium
abundance ($Y=0.4$; green), and flash-mixed models for stars
born at standard helium abundance (blue curves) and enhanced helium
abundance (brown curves).  
The combination of canonical and flash-mixed models is able to
reproduce the full luminosity width of the observed EHB, but not the
red colors seen in some of the BH stars.}
\end{figure}

The CMD for each cluster is shown in Figure 2.  Each cluster exhibits
a distinct hook feature at the hot end of its HB locus.  As discussed
extensively in \citet{brown_2001}, this hook feature cannot be due to
photometric scatter, differential reddening, or instrumental effects,
because these mechanisms could not affect the photometry of the EHB
without also affecting the photometry of the HB stars lying
immediately to the red of the EHB.  The EHB stars span
a much larger luminosity range, despite their similar far-UV
luminosities and photometric errors.

We compare our CMDs to stellar evolutionary models by transforming these
models to the observational plane using self-consistent 
non-LTE model atmospheres and
synthetic spectra \citep{brown_2010}.  We begin with
canonical evolutionary models at standard helium abundance ($Y=0.23$;
Figure 2).  We adopt metallicity and distance values that are
representative of values in the literature, and then make small
adjustments to the extinction so as to place the blue HB stars ($7000
\lesssim T_{\rm eff} \lesssim 15,000$~K) in the model at the expected
location.  For the four metal-poor clusters, the theoretical HB
distribution should be coincident with the observed HB locus at these
intermediate temperatures.  For the two metal-rich clusters, however, 
\citet{rich_1997} have shown that the HB slopes strongly upward as one moves
from the reddest HB stars to the top of the blue HB tail
\citep[see also][]{busso_2007},
such that the blue HB is $\sim$0.5~mag brighter than one
would expect for a canonical locus normalized to the red HB.  This
increase in HB luminosity at hotter temperatures is what one would
expect if the hotter HB stars were preferentially drawn from
populations increasing in helium abundance, up to $Y\sim 0.4$ on the EHB.
For this reason,
we normalize the canonical HB locus for these two clusters (Figure 2)
to a point 0.5~mag fainter than the observed blue HB.
Of course, it would be simpler to consistently align
our models to the red HB in all of our clusters, but it falls outside
of our UV CMDs, thus requiring us to use the blue HB.  

Next, in Figure 3, we compare our photometry to the expectations 
from evolutionary tracks with enhanced helium ($Y=0.4$),
assuming the same cluster parameters (reddening and distance) as for the
standard $Y$ tracks.  At first glance, an
enhanced helium abundance offers a promising solution to the BH
puzzle, because HB models with enhanced helium exhibit a sharp
downturn at the hot end of the HB, reminiscent of the BH
feature.  However, the downturn in such a model arises from the
combination of two effects: a relatively small decrease in luminosity
on the hot end of the HB, and a significant increase in luminosity at
cooler temperatures (compare violet and green models in Figure 3).
As is evident in Figure 3, the locus of the $Y=0.4$ models is much
brighter than the blue HB population in all of the clusters.
For this reason, a $Y=0.4$ HB model cannot, by itself, explain the observed
luminosity difference between the blue HB and BH stars.  However, because it is
easier to produce EHB stars in a helium-enriched population, the HB
population in each cluster may host EHB stars preferentially drawn
from that subset of the cluster population enhanced in helium.  The
existence of helium-enhanced subpopulations in massive globular
clusters, including NGC~2808 \citep{dantona_2005,piotto_2007}
and NGC~6715 \citep{layden_2000,siegel_2007}
is the leading explanation for the splitting of the main
sequence and subgiant branch recently discovered in optical CMDs of
these clusters.  In those clusters exhibiting multiple main sequences,
the bluer main sequence is attributed to a higher helium abundance
up to $Y \sim 0.4$ \citep{piotto_2005}; in $\omega$ Cen, the bluer 
main sequence is actually more metal rich.

Finally, we compare our CMDs to models for stars that underwent a late
helium-core flash on the white dwarf cooling curve (Figure 3).  We
show such models for stars born on the main sequence with both
standard helium abundance ($Y=0.23$; blue curves) and enhanced helium
abundance ($Y=0.4$; brown curves).  
Instead of plotting the core helium-burning locus of all BH stars, 
we show three representative tracks in each
category; two of these tracks fall at the extremes of the range in the
mass loss rate leading to flash mixing, while one falls halfway
between these extremes.  Although the range of mass loss that leads to
flash mixing is relatively large, all of the flash-mixed tracks
produced by such flash mixing fall within a narrow temperature range.
The offset in luminosity among these tracks is due to small
differences in the helium-core mass.  As can be seen in Figure 3,
flash-mixed models can reproduce the faint luminosities observed in
the BH population, particularly the flash-mixed models originating in
a subpopulation with enhanced helium.

There is one obvious problem, however.  
Flash-mixed models (whether originating in a population with enhanced
or standard $Y$) cannot reproduce the reddest stars in the BH population.
\citet{brown_2001} explored various ways of producing such red
subluminous stars, including enhanced line blanketing in the
atmosphere, due to an increase in the atmospheric iron abundance via
radiative diffusion.  This explanation is explored in more detail
in the next section, where we obtained spectroscopy of two of these
curiously red BH stars.

The BH phenomenon appears to be restricted to the most massive
globular clusters in the Galaxy \citep[and references
  therein]{brown_2010}. Of the clusters that have extended HBs and UV
imaging, only those clusters with masses exceeding 1.2$\times$10$^6$
M$_\odot$ exhibit BH stars.  This is not because massive
clusters offer more stellar mass for finding rare objects.  If one
constructs a virtual globular cluster by summing all of the known
low-mass clusters that have extended HBs and UV imaging, this virtual
cluster would have a mass of 5.1$\times$10$^6$ M$_\odot$ (exceeding
any known globular cluster), yet it would host only a few BH
stars; we would expect dozens of such stars if we were to scale from
the numbers of BH stars present in actual massive clusters.
One can also see the importance of cluster mass by comparing the UV
CMDs of individual clusters hosting similar numbers of EHB stars.  For
example, the published CMDs of NGC~6752 \citep[$M \approx
  3$$\times$10$^5$ $M_\odot$]{landsman_1996} and NGC~2808 \citep[$M \approx
3$$\times$10$^6$ $M_\odot$]{brown_2001} each include approximately 80 EHB stars,
but all of the EHB stars in NGC~6752 are normal, while half of the EHB
stars in NGC~2808 are subluminous (i.e., BH stars).  Because a
delayed core flash is more likely in stars that were born with an
enhanced helium abundance ($Y \sim 0.4$), BH stars may only
form in those clusters massive enough to retain the helium-rich ejecta
from the first stellar generation of stars,

\section{Spectroscopy}

Spectroscopic evidence for flash mixing includes 
{\it Far Ultraviolet Spectroscopic Explorer (FUSE)}
observations of field stars,
Very Large Telescope (VLT) observations
of EHB stars in $\omega$ Cen, and
{\it HST}/STIS observations of
both normal and subluminous EHB stars in NGC~2808.

\begin{figure}[t!]
\plotone{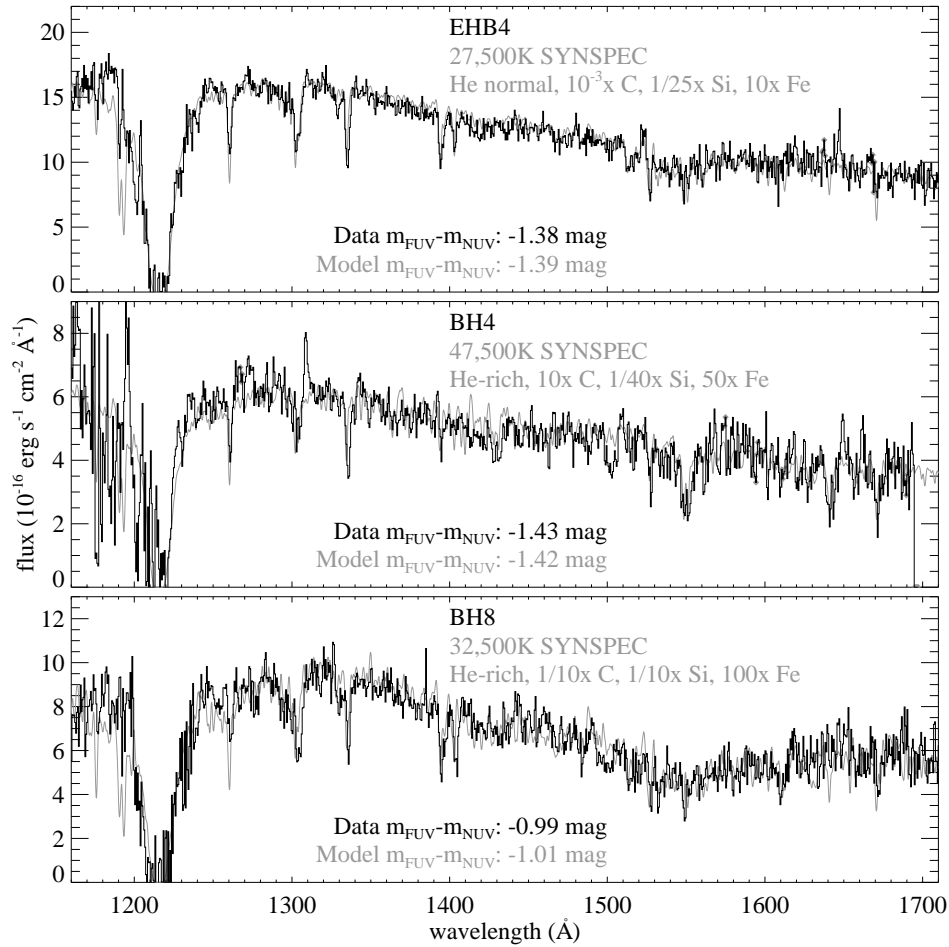}
\caption{Representative spectra from our sample: a
  normal EHB star ({\it top panel}), a BH star ({\it middle panel}),
  and one of the unusually red subluminous stars ({\it bottom panel}).
  In each panel, we also show a synthetic spectrum (grey curve) that
  matches both the observed spectrum \citep{brown_2012} and UV
  photometry \citep{brown_2001}. Compared to the normal EHB stars, the
  BH stars are hotter and have stronger helium and carbon lines, as
  expected if the BH stars formed via flash mixing.  The BH stars all
  exhibit enhanced iron-peak abundances, while the normal EHB stars
  are mixed, with some enhanced and others not.  Some subluminous EHB
  stars are much redder than expected (see Figure 3), but their red
  colors can be explained by a large enhancement of iron-peak
  elements, presumably from radiative levitation.}
\end{figure}

In the Galactic field population, sdB stars are the analogs to the EHB
stars found in globular clusters.  Because the helium-rich sdB stars (He-sdB
stars) seemed to be the most likely candidates for flash mixing in the
field population, \citet{lanz_2004} obtained {\it FUSE} spectroscopy of
three He-sdB stars to characterize their helium, carbon, and nitrogen
abundances.  Two of these stars (PG1544+488 and JL87) exhibit enormous
carbon enhancements, in excellent agreement with the expectations from
flash mixing.  Specifically, PG1544+488 has a surface composition of
96\% helium, 2\% carbon, and 1\% nitrogen, while JL87 has a similar
composition (although it retains more surface hydrogen).  The spectra
of these two stars are among the strangest in the {\it FUSE} archive, with
carbon lines stronger than the Lyman series.

$\omega$ Cen is the most massive globular cluster in the Galaxy, and
the cluster where UV photometry first revealed the existence of BH
stars \citep{dcruz_1996,dcruz_2000}.  Using optical photometry of
the cluster, \citet{moehler_2011} selected the faintest and hottest
EHB stars in $\omega$ Cen to produce a sample that was likely to
include BH stars, although these stars did not have the UV
photometry required to demonstrate that any particular star is subluminous.
They obtained FLAMES+GIRAFFE spectroscopy on the VLT over the spectral
range of 3964--4567~\AA\ for 109 EHB stars.  They found that the stars
at $T_{\rm eff} <$30,000~K are helium-poor, but that nearly three quarters
of the stars at hotter temperatures have solar or enhanced helium
abundance.  They also found the carbon abundance to be strongly correlated
with the helium abundance, with carbon approaching $\sim$3\% by mass.

In late 2010 and early 2011, we obtained {\it HST}/STIS spectroscopy
of 24 hot stars in NGC~2808, drawn from its UV CMD 
\citep[Figure 3]{brown_2001}. This sample includes 7 normal EHB stars and 8
BH stars.  It also includes 5 cooler
stars on the blue HB, a post-EHB star, and three unclassified objects
significantly hotter than the canonical EHB.  Here we restrict the
discussion to the 15 EHB and BH stars \citep[for the full sample, see][]
{brown_2012}.

Figure 4 shows representative spectra from our sample.
For comparison, we show a non-LTE
synthetic spectrum that matches the far-UV spectroscopy \citep{brown_2012}
and UV photometry \citep{brown_2001}.  To ease
interpretation, the abundances of carbon, silicon, and iron are
specified relative to the mean cluster abundance, which we assume to
be [Fe/H]=~$-1.36$ with [$\alpha$/Fe]=0.3.  Our ``helium normal''
models assume $Y=0.23$ while our ``helium rich'' models assume
$Y=0.99$.  Figure 5 shows the composite spectrum of all normal EHB
stars in our sample compared to the composite spectrum of all BH
stars in our sample.  The BH stars are much hotter than
the normal EHB stars, and they exhibit much stronger carbon and helium
features, as expected if the BH stars formed via flash mixing.
The silicon lines vary significantly from star to star, presumably due
to atmospheric diffusion, but on average the normal EHB stars and the
BH stars have similar silicon abundances.  
Two of our BH stars fall in the temperature range (48,000~K~$\lesssim
T_{\rm eff} \lesssim$~52,000~K) spanned by pulsating subdwarfs
recently discovered in $\omega$~Cen \citep{randall_2011}.
\citet{randall_2011} suggest this is a new instability strip, and so
our hottest BH stars may also be pulsators.  Such an instability strip
would be better populated in massive globular clusters than in 
the Galactic field population; massive clusters host
subpopulations enhanced in helium ($Y\sim 0.4$), and such populations
are more likely to produce hotter subdwarfs through flash mixing.

All of the BH stars exhibit enhanced iron-peak abundances, while the
normal EHB stars are mixed, with some exhibiting normal iron-peak
abundances and others showing enhancement.  The distortion of the
normal abundance pattern by atmospheric diffusion in hot subdwarfs is
well documented \citep[e.g.,][]{heber_2009}, although other processes,
such as turbulent mixing and mass loss, also play a role in these
anomalies \citep{hu_2011}.  The two reddest stars in our BH sample
exhibit deep and broad absorption troughs from the iron-peak elements
(Figure 4).  Subluminous EHB stars with such red colors have been
difficult to explain \citep[Figure 3]{brown_2010}, but our spectra
indicate that they may be extreme cases of metal enhancement through
radiative levitation.

\begin{figure}[t!]
\plotone{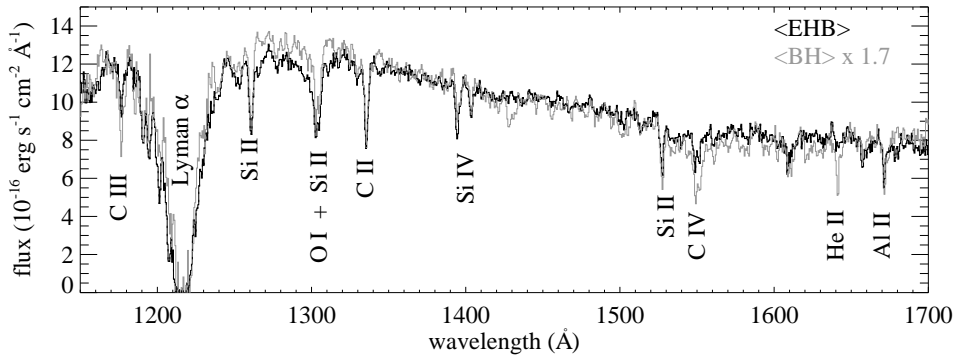}
\caption{The composite spectrum for all normal EHB stars in our sample
(black curve) compared to the composite spectrum for all BH stars
in our sample (grey curve).  The Si II, O I, C II, and All II lines are
all interstellar.  Si IV and C IV have both an interstellar component
and a photospheric component.  C III and He II are purely photospheric.
The BH spectrum has stronger carbon and helium absorption, as 
expected from flash mixing.}
\end{figure}

The fact that the BH stars show much stronger helium and carbon
lines, when compared to the normal EHB stars, is qualitatively in line
with expectations from flash mixing.  However, in the BH
sample, the carbon abundance is not as high as one would expect for a
star that recently emerged from flash mixing.  In a flash-mixed star,
carbon (and/or nitrogen) should comprise up to 4\% of the atmosphere
by mass \citep[see, for example,][]{lanz_2004}.  In contrast, the
strongest carbon enhancement found in our BH sample corresponds
to a carbon abundance that is 0.2\% by mass.  Furthermore, the carbon
abundance in the normal EHB sample is strongly reduced with respect to
the mean abundance in the cluster, and is orders of magnitude smaller than
in the BH stars.  It appears that both the normal
EHB sample and the BH sample suffer from depleted carbon
abundances.  This is not unexpected.  The numerical simulations of
\citet{miller_2008} trace the atmospheric abundances of
stars as they evolve through the flash mixing stage and into the
subsequent period of stable core helium burning, accounting for
atmospheric diffusion due to gravitational settling and radiative
levitation.  They found that the carbon abundance declines rapidly
after a star emerges from the flash-mixing process, dropping by an
order of magnitude after 1000 years, and by several orders of
magnitude 10$^7$ years later.  The star spends only a small fraction
of its stable core helium-burning lifetime with a carbon abundance
near its maximum of $\sim$1--4\% by mass.  The diffusion models indicate
that the helium abundance will also decline with time.  However, these results
assume that atmospheric diffusion is not inhibited by such processes as
mass loss, turbulence, and surface convection.  

\section{Summary}

Although blue hook stars are rare, they have a significant presence in
the UV CMDs of massive globular clusters.  Such clusters also possess
complex optical CMDs, which exhibit splitting of the main sequence and
subgiant branch, indicating the presence of subpopulations enhanced in
helium \citep[$Y \sim 0.4$;][]{piotto_2005}.  Although stars born with
enhanced helium ($Y \sim 0.4$) are more likely to become EHB stars, we
have shown that only flash mixing leads to the luminosity range needed
to produce the blue hook feature in the UV CMDs of massive globular
clusters.  For those stars suffering high mass loss on the RGB, the
helium-core flash will be delayed to the WD cooling curve. The
flash mixing of the envelope associated with such a delayed helium-core
flash will produce hotter EHB stars that
are enhanced in atmospheric helium and carbon, relative to their
normal EHB brethren.  The larger bolometric correction and reduced
opacity below the Lyman limit in the flash-mixed stars makes them
appear subluminous in the UV
and optical.  Our UV spectroscopy of both normal and subluminous EHB
stars in one such cluster (NGC~2808) demonstrates that the blue hook
population is hotter than the normal EHB population, even though the normal
and subluminous EHB samples span similar ranges in UV color.  Moreover,
the blue hook stars exhibit much stronger absorption lines of carbon
and helium, as expected from flash mixing.  It is also clear that
both the EHB and BH samples have abundance patterns that have been
altered by atmospheric diffusion.  The reddest of the subluminous EHB
stars may be explained by radiative levitation of the iron-peak elements
in the atmospheres of these stars.

\acknowledgements
Support for programs 10815 and 11665 are provided through a NASA grant from
STScI, which is operated by AURA under contract NAS 5-26555.

\vspace{-0.08in}

\bibliography{tbrown_sdob5}

\end{document}